\documentclass[twocolumn,twoside,pra,aps,superscriptaddress,showpacs]{revtex4}
\usepackage{amssymb}

\usepackage{amsmath,mathrsfs,amsbsy,color,graphicx,bm,amsthm,amsfonts}
\usepackage{units}
\usepackage{bbm}

\begin{document}

\title{Tunable single-photon diode by chiral quantum physics}
\author{Wei-Bin Yan }
\affiliation{Liaoning Key Lab of Optoelectronic Films \& Materials, School of Physics and
Materials Engineering, Dalian Nationalities University, Dalian, 116600, China}
\author{Wei-Yuan Ni}
\affiliation{School of Science, Changchun University of Science and Technology,
Changchun, 130022, China}
\author{Jing Zhang}
\affiliation{Liaoning Key Lab of Optoelectronic Films \& Materials, School of Physics and
Materials Engineering, Dalian Nationalities University, Dalian, 116600, China}
\author{Feng-Yang Zhang}
\email{zhangfy@dlnu.edu.cn}
\affiliation{Liaoning Key Lab of Optoelectronic Films \& Materials, School of Physics and
Materials Engineering, Dalian Nationalities University, Dalian, 116600, China}
\author{Heng Fan}
\email{hfan@iphy.ac.cn}
\affiliation{Beijing National Laboratory for Condensed Matter Physics, Institute of
Physics, Chinese Academy of Sciences, Beijing 100190, China}

\begin{abstract}
We investigate the single photon scattering by an emitter chirally
coupled to a one-dimensional waveguide. The single-photon transport
property is essentially different from the symmetrical coupling
case. The single photons propagating towards the emitter in opposite
directions show different transmission behaviors, which is a
manifestation of the single-photon diode. In the ideal chiral
coupling case, the transmission probability of the single photon
transport in one direction is zero by critical coupling, while in
the opposite direction it is unity. The diode works well only when
the single-photon frequency meets certain conditions. For a
two-level emitter, the diode works well when the single photon is
nearly resonant to the emitter. For a $\Lambda $-type three-level
emitter, when the single-photon frequency is greatly altered, we can
adjust the parameters of the external laser to ensure the diode
works well. The latter provides a manner to realize a single-photon
switch, in which the single-photon transmission probability can
reach zero or unity although the emitter's decay is considered.
\end{abstract}

\maketitle


\bigskip







\section{Introduction}

Optical diode, which allows unidirectional propagation of light, requires
the ability to break Lorentz reciprocity\cite{lobroken}. Nonreciprocity in
light propagation has been extensively studied by various physical
mechanisms \cite%
{nrgallo,nrlira,nrzaman,nr1bi,nr2fan1,nr3kim,nrkim,nr4kamal,nr5sounas,nr6roy, nr6roy1%
,nr8estep,nr9fra,sinshenh,nr12fra,nr10yu,nr11shi,nr13miro,nr14fan,nr15bender,chemt, nr16rod%
,nr17peng,nr18chang,sinshen,sinlenf,sinxia,sinsayrin,sinxu,chir1mitsch, chir2petersen,chir3feber,sinyan,chmttt,chir4}%
. Single photons are considered as the ideal carrier of quantum
information. The single-photon optical diode with low losses is an
indispensable element for future quantum networks\cite{netkimble}
requisite for optical quantum information and quantum computation.
Recently, the
single-photon diode has been successfully achieved, such as \cite%
{sinshen,sinlenf,sinxia,sinshenh,sinsayrin,sinxu}. These diodes work
well at a given frequency. If the input frequency is greatly
altered, the devices should be programmed and actively reconfigured\cite%
{sinsayrin}. Therefore, the largely tunable single-photon diode still needs
to be explored.

For this purpose, we propose a scheme to realize a largely tunable
single-photon diode. Our nonreciprocal system is realized by chiral
quantum optics
\cite{sinsayrin,chir2petersen,chir3feber,chiralop,chir4}. In chiral
quantum optics, the light propagating towards opposite directions
could be coupled to the emitter with different strengths. The chiral
coupling is underpinned by the spin-momentum locking of the
transversely confined light and the polarization dependent dipole
transitions of the emitter. In our scheme, the photon is largely
confined in a one-dimensional(1D) waveguide, which is chirally
coupled to an emitter. The photon scattering in 1D waveguide
symmetrically coupled to emitters has
been extensively studied\cite%
{Shen2005ol,Lan,Shen2007pra,royb,Fan2012prl,kocabas,111,Shenjt,fanshen,Paolo,busch,Shit,Zhengprl,Zhengprl1,Zhengprl2,Liao,Gongzr,Liaojq,Weilf1,Cheng,Wangzh,%
a11,a12,a13,a14,a15,a16,a17,Witthaut2010njp,a1,liqiong,Hart,Weilf,zhour,brad,a2,Yanwb,fengzb,liwenan,yanwb1,yanwb2,liaozy,liaozy1,fong,wuxw,chenmt,chengmt,chenggg,songgz,qinw,lity,shit11}%
. In the quantum network, the waveguide can act as a channel, and
the emitter as a nod. In the chirally coupling case, the
single-photon shows essentially different transport
properties\cite{sinxia,chemt,sinsayrin,sinyan,chmttt} compared to
the symmetrical coupling cases. We theoretically study the
single-photon scattering in the 1D waveguide chirally coupled to a
two-level emitter and a $\Lambda $-type three-level emitter,
respectively. When the decays from the emitter's excitation to the
other channels except the waveguide are neglected, the
nonreciprocity in single-photon propagation can not be achieved. The
single-photon reflection probabilities can not reach unity due to
the chiral coupling. When the decays are considered, the
transmission probabilities of the single photons transporting
towards the opposite directions are not equal. Under certain
conditions, the transmission probability for one of the directions
is zero due to the critical coupling, while the transmission
probability for the other direction reaches a near unity value. In
the ideal chiral coupling case, the emitter is decoupled to the
single photon transporting in one of the directions. The single
photon transporting in one direction completely decays to the other
channels except waveguide, while the single photon transporting in
the other direction will be completely transmitted due to the
decoupling. For the scheme composed by a 1D waveguide chirally
coupled to a $\Lambda $-type three-level emitter, an external laser
is employed to drive the emitter. It is significant that the
single-photon diode works well at different frequencies by
programming and actively adjusting the laser parameters.

Our scheme also shows certain advantages of the single-photon
switch. The control of single-photon transport in 1D waveguide has
been extensively investigated, such as
\cite{Shen2005ol,Lan,Weilf,Yanwb,Witthaut2010njp}. It is known that
when the emitter's decay is neglected, the single-photon
transmission probability can be zero or unity. However, when the
emitter's decay is considered, this perfect outcome can not be
realized. Having considered the emitter's decay, the transmission
probability of the single photon, which transports in the 1D
waveguide chirally coupled to a $\Lambda $-type three-level emitter,
can be zero by critical coupling or be unity by EIT
(Electromagnetically Induced Transparency).

\section{Model and single-photon scattering}

\begin{figure}[h]
\includegraphics*[width=6cm, height=4cm, bb=125 211 575 492]{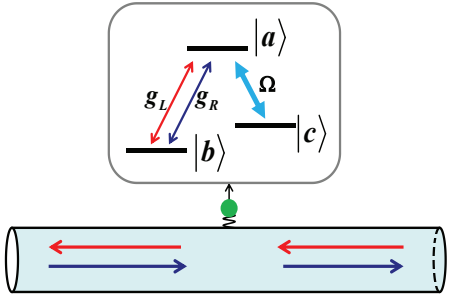}
\caption{Schematic configuration of the tunable single-photon diode.
A $\Lambda $-type three-level emitter driven by an external laser is
chirally coupled to a 1D waveguide. }
\end{figure}

We consider a $\Lambda $-type three-level emitter chirally coupled to a 1D
waveguide. The emitter's states are denoted by $\left| b\right\rangle $, $%
\left| a\right\rangle $ and $\left| c\right\rangle $, with the
corresponding level frequencies $\omega _{m}$ $(m=a,b,c)$,
respectively. The ground-state energy is for reference so that
$\omega _{b}$ is taken to be zero. The right-moving and left-moving
photons in the waveguide are coupled to the level transition $\left|
b\right\rangle \longleftrightarrow \left| a\right\rangle $ with
strengths $g_{R}$ and $g_{L}$, respectively. The coupling strengths
are assumed to be independent of the photon wave number, which is
equivalent to the Weisskopf-Wigner approximation. The coupling
strengths $g_{R}$ and $g_{L}$ are not equal, distinguishing from the
symmetrical coupling case. For simplicity and without generality, we assume $%
\Gamma _{R}\geqslant \Gamma _{L}$ in this paper. In the ideal
chirally coupling case, the photon is perfectly circularly polarized
at the position of the emitter and hence the polarization is
orthogonal for opposite propagation. Consequently, the level
transition $\left| b\right\rangle \longleftrightarrow \left|
a\right\rangle $ can be solely coupled to one propagation direction
photon. To measure the chiral coupling character, we
bring in the parameter $C=\left| \frac{ \Gamma _{R}-\Gamma _{L} }{%
 \Gamma _{R}+\Gamma _{L}}\right|$. Obviously, $C=0$
when $\Gamma
_{R}=\Gamma _{L}$, $0<C\leqslant 1$ when $\Gamma _{R}\neq \Gamma _{L}$, and $%
C=1$ in the ideal chiral coupling case. We employ an external laser
beam with frequency $\omega _{L}$ and Rabi frequency $\Omega $ to
drive the level transition $\left| c\right\rangle
\longleftrightarrow \left| a\right\rangle $ by frequency selection.
When $\Omega \neq 0$, the system can be considered as a 1D waveguide
chirally coupled to a dressed three-level emitter. Within the
rotating wave approximation, the time-independent
Hamiltonian governed by the system is%
\begin{eqnarray}
H &=&-i\int dxa_{R}^{\dagger }(x)a_{R}(x)+i\int dxa_{L}^{\dagger }(x)a_{L}(x)
\notag \\
&&+(\omega _{a}-i\frac{\gamma _{a}}{2})\sigma ^{aa}+(\omega _{c}+\omega
_{L})\sigma ^{cc}  \notag \\
&&+[g_{R}\int dx\delta (x)a_{R}(x)\sigma ^{ab}+g_{L}\int dx\delta
(x)a_{L}(x)\sigma ^{ab}  \notag \\
&&+\Omega \sigma ^{ac}+h.c.]\text{,}  \label{hami1}
\end{eqnarray}%
with $\sigma ^{mn}=\left| m\right\rangle \left\langle n\right| $ ($m,n=a,b,c$%
) being the raising, lowering and energy level population operators
of the emitter. The operators $a_{R}^{\dagger }(x)$ and
$a_{L}^{\dagger }(x)$ create a right-moving and left-moving photons
in the waveguide at the site $x$, respectively. The parameter
$\gamma _{a}$ accounts for the loss from the emitter's excitation
to\ the other channels except the waveguide, such as the spontaneous
emission to the free space. We have taken $\hbar =1$, and
the photonic group velocity $v_{g}=1$. The first line of the Hamiltonian (%
\ref{hami1}) denotes the free part of the waveguide photon. The
second line is the emitter's energy including the intrinsic
dissipation, which is represented by adding the imaginary part
$-i\frac{\gamma _{a}}{2}$ to the corresponding level energy in the
quantum jump picture. Here we assume that the states $\left|
b\right\rangle $ and $\left| c\right\rangle $ are long-live states,
and $\left| a\right\rangle $ is the excited state. The third line
represents the coupling of the waveguide photon to the emitter. In
this paper, initially, the emitter is in the state $\left|
b\right\rangle $ and a single photon is injected into the waveguide
from left or right side. The frequency of photon transporting in the
waveguide is far away from the cutoff frequency of the waveguide so
that the photonic dispersion relation is approximately linearized.

The system scattering eigenstate has the form of%
\begin{eqnarray}
\left\vert \Psi \right\rangle &=&\int dx[\alpha _{R}(x)a_{R}^{\dagger
}(x)+\alpha _{L}(x)a_{L}^{\dagger }(x)]  \notag \\
&&+\beta _{a}\sigma ^{ab}+\beta _{c}\sigma ^{cb}]\left\vert \phi
\right\rangle \text{,}  \label{state1}
\end{eqnarray}%
with $\alpha _{R}(x)$, $\alpha _{L}(x)$, $\beta _{a}$ and $\beta _{c}$ being
probability amplitudes. The state $\left\vert \phi \right\rangle $ denotes
that the emitter is in its ground sate $\left\vert b\right\rangle $ and the
number of the photon transporting in the waveguide is zero. The probability
amplitudes can be obtained from the eigenequation $H\left\vert \Psi
\right\rangle =E\left\vert \Psi \right\rangle $, with eigenvalue $%
E=v_{g}\left\vert k\right\vert $.

When the input photon is injected from the left side of the waveguide, the
spatial dependence of the amplitudes $\alpha _{R}(x)$ and $\alpha _{L}(x)$
are taken as $\alpha _{R}(x)=[\theta (-x)+t_{R}\theta (x)]e^{ikx}$ and $%
\alpha _{L}(x)=r_{R}\theta (-x)e^{-ikx}$, respectively. The function $\theta
(x)$ is the Heaviside step function. The parameters $t_{R}$ and $r_{R}$
represent the single-photon transmission and reflection probability
amplitudes, respectively. The subscript R in $t_{R}$ and $r_{R}$ denotes
that the input photon is right moving. The expressions of $t_{R}$ and $r_{R}$
are%
\begin{eqnarray}
t_{R} &=&\frac{\Delta _{k}(\delta _{k}-i\frac{\gamma _{a}}{2}+i\frac{\Gamma
_{R}-\Gamma _{L}}{2})+\Omega ^{2}}{\Delta _{k}(\delta _{k}-i\frac{\gamma _{a}%
}{2}-i\frac{\Gamma _{R}+\Gamma _{L}}{2})+\Omega ^{2}}  \notag \\
r_{R} &=&\frac{i\sqrt{\Gamma _{R}\Gamma _{L}}\Delta _{k}}{\Delta _{k}(\delta
_{k}-i\frac{\gamma _{a}}{2}-i\frac{\Gamma _{R}+\Gamma _{L}}{2})+\Omega ^{2}}
\label{ampl1}
\end{eqnarray}%
where $\delta _{k}=\omega _{ab}-v_{g}\left| k\right| $ and $\Delta
_{k}=\Delta -\delta _{k}$ are detunings, with $\Delta =\omega _{ac}-\omega
_{L}$. The parameters $\Gamma _{R}=\frac{g_{R}^{2}}{v_{g}}$ and $\Gamma _{L}=%
\frac{g_{L}^{2}}{v_{g}}$ account for the spontaneous emissions from the
emitter's excitation into the waveguide right-moving and left-moving
channels, respectively.

Similarly, when the input photon is injected from the right side, the
single-photon transmission amplitude $t_{L}$ and reflection amplitude $r_{L}$
are%
\begin{eqnarray}
t_{L} &=&\frac{\Delta _{k}(\delta _{k}-i\frac{\gamma _{a}}{2}+i\frac{\Gamma
_{L}-\Gamma _{R}}{2})+\Omega ^{2}}{\Delta _{k}(\delta _{k}-i\frac{\gamma _{a}%
}{2}-i\frac{\Gamma _{R}+\Gamma _{L}}{2})+\Omega ^{2}}  \notag \\
r_{L} &=&r_{R}\text{.}  \label{ampl2}
\end{eqnarray}%
If the decay rates from the emitter' excitation to the other channels are
neglected, i. e., $\gamma _{a}=0$, the single-photon transmission
probabilities $T_{R}=\left| t_{R}\right| ^{2}$ and $T_{L}=\left|
t_{L}\right| ^{2}$ are equal. The single-photon diode can not be achieved
although the emitter are chirally coupled to the 1D waveguide. However, when
the decay rate $\gamma _{a}$ is not negligible, it is interesting that $%
T_{R} $ and $T_{L}$ are different from each other due to the chiral
coupling. The reflection probabilities $R_{R}=\left| r_{R}\right| ^{2}$ and $%
R_{L}=\left| r_{L}\right| ^{2}$ are equal in any case. For simplicity, we
label $R=R_{R}=R_{L}$. When $\Gamma _{R}=\Gamma _{L}$, our results agree
with the outcomes derived in the symmetrical coupling case \cite%
{Witthaut2010njp}.

\begin{figure}[t]
\includegraphics*[width=4cm, height=2.5cm]{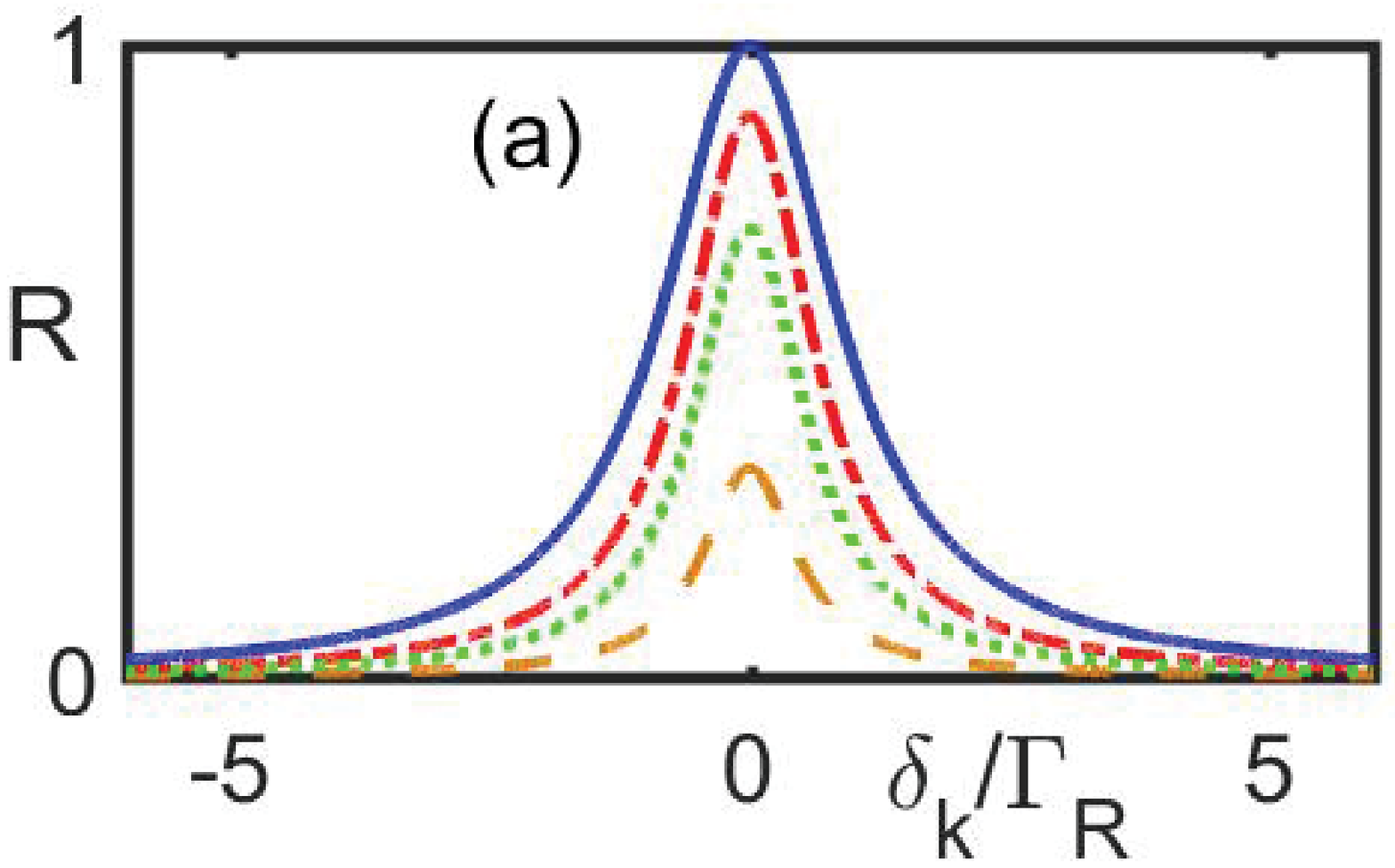} \includegraphics*%
[width=4cm, height=2.5cm]{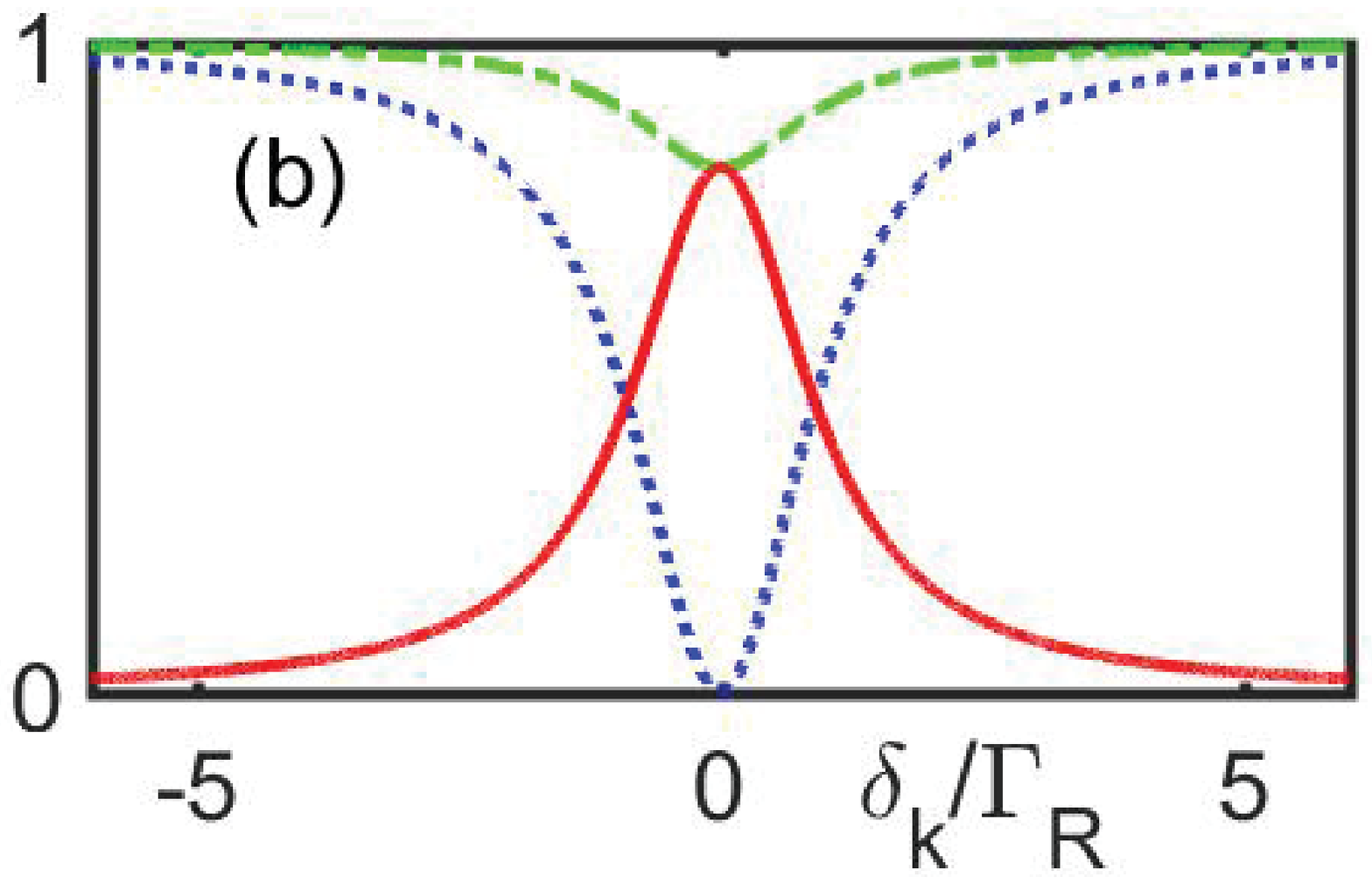} \includegraphics*[width=4cm,
height=2.5cm]{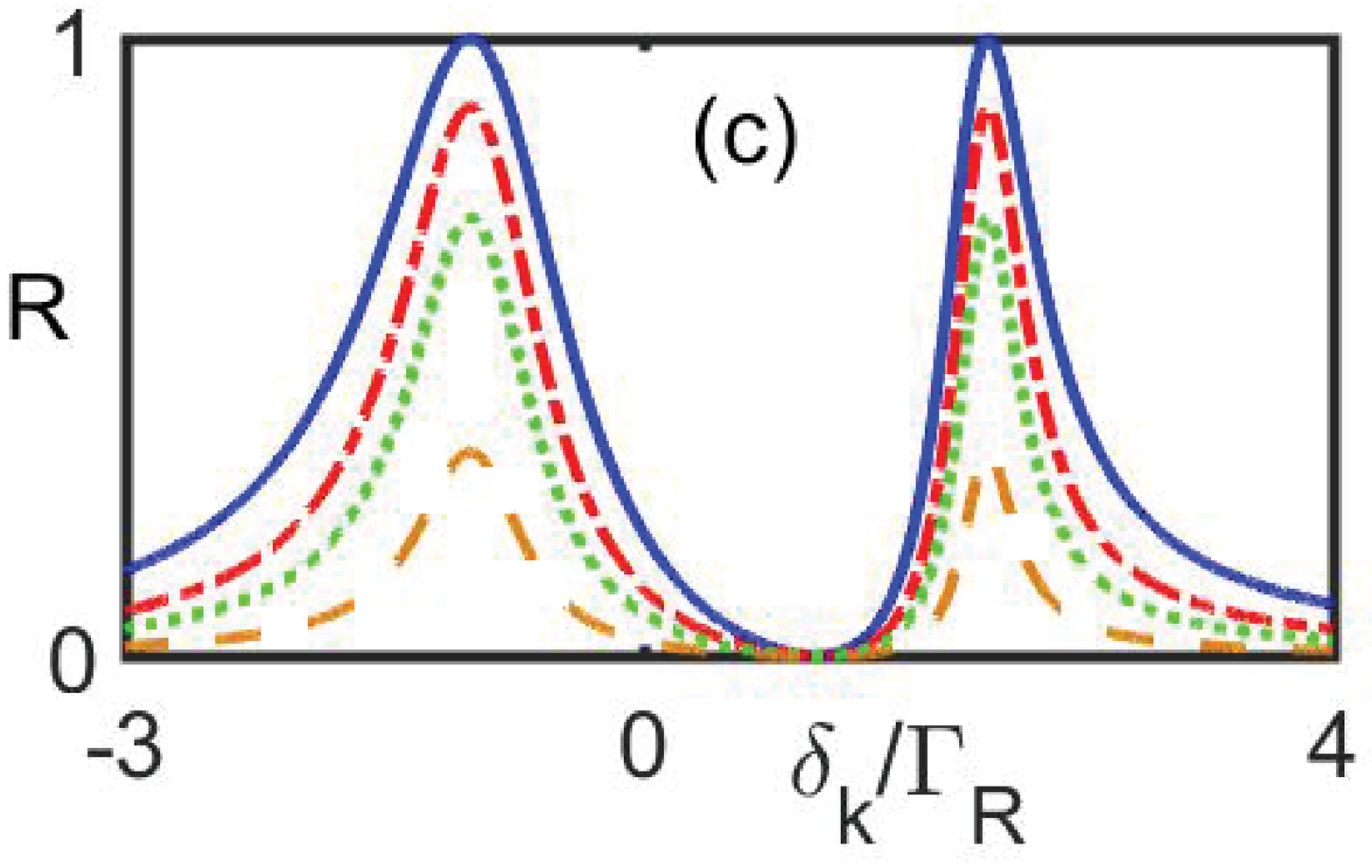} \includegraphics*[width=4cm, height=2.5cm]{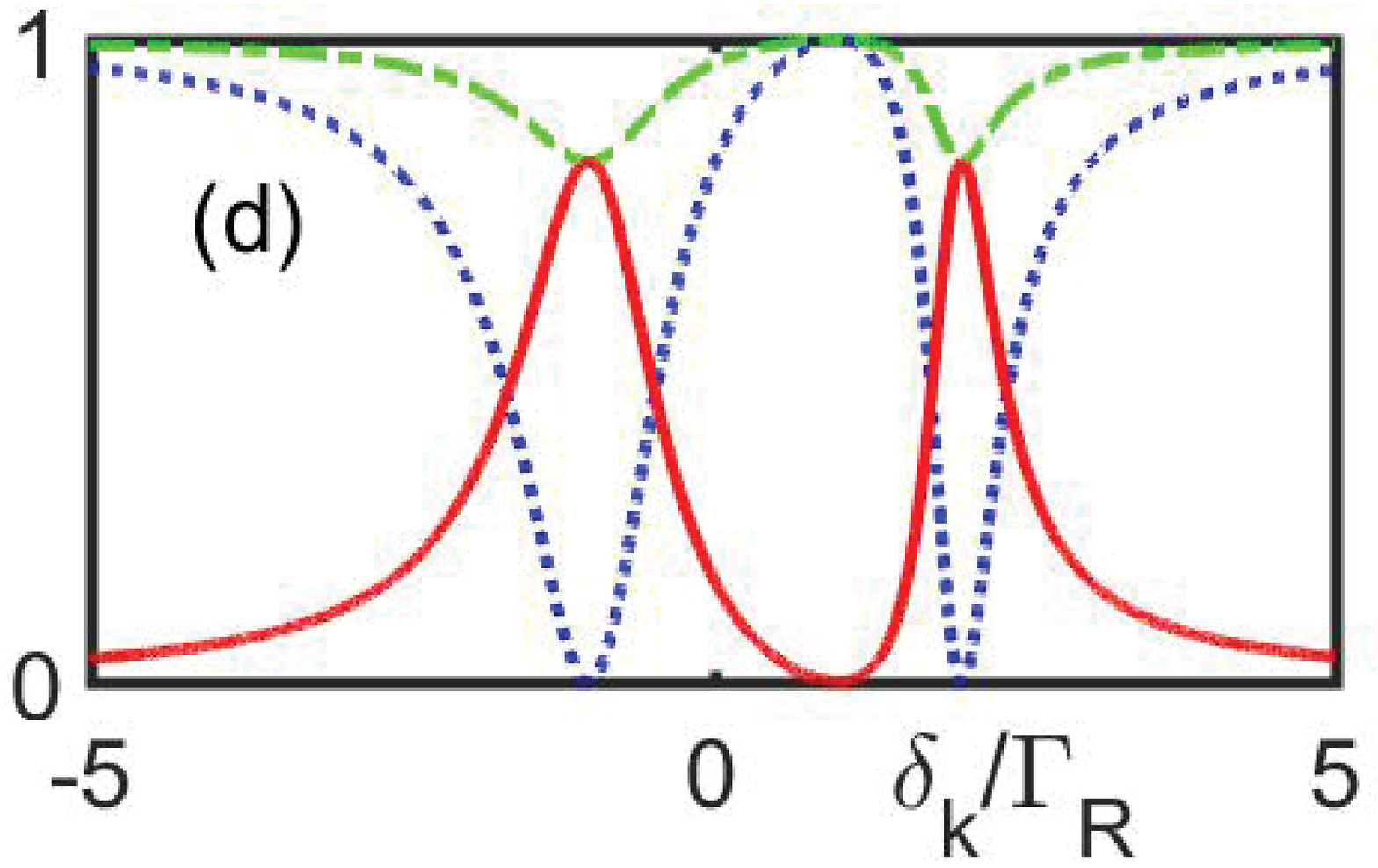%
}
\caption{Single-photon transport properties against the detuning $\protect%
\delta_k$. (a) and (c) show the reflection probability $R$ when $\protect%
\gamma_a=0$. (b) and (d) show the transmission probabilities $T_R$, $T_L$
and $\Delta T$ when $\protect\gamma_a=\Gamma_R-\Gamma_L$. (a) and (b)
correspond to the results of a two-level emitter coupled to the 1D
waveguide, (c) and (d) correspond to a $\Lambda$ type three-level emitter.
The orange dashed lines, green dotted lines, red dashed dotted lines, and
blue solid lines in (a) and (c) denote the situations $\Gamma_L/\Gamma_R=0.1$%
, $\Gamma_L/\Gamma_R=0.3$, $\Gamma_L/\Gamma_R= 0.5$, and $%
\Gamma_L/\Gamma_R=1 $, respectively. The blue dotted lines, green dashed
dotted lines, and red solid lines in (b) and (d) denote the probabilities $%
T_R $, $T_L$ and $\Delta T $ when $\Gamma_L/\Gamma_R=0.1$, respectively. The
parameters in (c) and (d) are $\Omega/\Gamma_R=2$, $\Delta/\Gamma_R=1$.}
\end{figure}

\section{Single-photon diode}

If the external laser is shut off, the emitter's level $\left|
c\right\rangle $ never participates in the dynamic process.
Consequently, our scheme is a 1D waveguide coupled to a two-level
emitter. In this case, the single photon transport properties are
\begin{eqnarray}
t_{R} &=&\frac{\delta _{k}-i\frac{\gamma _{a}}{2}+i\frac{\Gamma _{R}-\Gamma
_{L}}{2}}{\delta _{k}-i\frac{\gamma _{a}}{2}-i\frac{\Gamma _{R}+\Gamma _{L}}{%
2}}  \notag \\
t_{L} &=&\frac{\delta _{k}-i\frac{\gamma _{a}}{2}+i\frac{\Gamma _{L}-\Gamma
_{R}}{2}}{\delta _{k}-i\frac{\gamma _{a}}{2}-i\frac{\Gamma _{R}+\Gamma _{L}}{%
2}}  \notag \\
r_{R} &=&r_{L}=\frac{i\sqrt{\Gamma _{R}\Gamma _{L}}}{\delta _{k}-i\frac{%
\gamma _{a}}{2}-i\frac{\Gamma _{R}+\Gamma _{L}}{2}}\text{.}  \label{ampl3}
\end{eqnarray}%
In the symmetrical coupling case, i. e., $\Gamma _{R}=\Gamma _{L}$,
the single-photon transport in a waveguide coupled to a two-level
emitter has been extensively studied. It is known that when the
emitter's decay to other modes except waveguide mode is neglected,
the single photon moving towards the
emitter will be fully reflected by interference in the resonance case\cite%
{Shen2005ol}. In the chiral coupling case, the input single photon
can not be fully reflected in any case. From the expression
(\ref{ampl3}), the maximum value of $R$ is obtained as $1-C^{2}$ in
the resonance case. In Fig. 2(a), we plot the single photon
reflection probabilities against the detuning $\delta _{k}$ for a
two-level emitter coupled to a 1D waveguide. The spectra are shaped
like the Lorentzian line. As the value of $C$ decreases, the maximum
value of $R$ increases. This can also be understood from the fact
that the input right-moving (left-moving) photon is converted into
left-moving (right-moving) photon by the emitter-waveguide
interaction. The reflection probability $R$ is essentially the
conversion efficiency. From the investigations in
\cite{brad,yanwb1}, the conversion efficiency, which
relates to the difference between the coupling strengths $g_{R}$ and $g_{L}$%
, can reach unity only in the symmetrical coupling case.

For the two-level emitter, the critical coupling condition can not
be satisfied for any nonzero value of $\gamma _{a}$ in the
symmetrical coupling case. In the chiral coupling case, when $\delta
_{k}=0$ and $\gamma _{a}=\Gamma _{R}-\Gamma _{L}$, we obtain
$T_{R}=0$, $T_{L}=(\frac{\Gamma
_{R}-\Gamma _{L}}{\Gamma _{R}})^{2}$, and $R_{R}=R_{L}=\frac{\Gamma _{L}}{%
\Gamma _{R}}$. The resonant single photon injected into the left
port of the waveguide can not be received from the right port due to
the critical coupling. However, the single photon injected into the
right port will be received from the left port with a near unity
probability when $C\rightarrow 1$. In the ideal chiral coupling
case, the photon injected from the left side will completely decay
out of the waveguide, i. e. $R=0$. The photon injected from the
right hand will be completely transmitted because it is decoupled to
the emitter. The difference between the transmission probabilities
corresponding to opposite transport directions is $\Delta
T=\left\vert T_{R}-T_{L}\right\vert =\frac{\gamma _{a}(\Gamma
_{R}-\Gamma _{L})}{\delta _{k}^{2}+(\frac{\gamma _{a}+\Gamma
_{R}+\Gamma _{L}}{2})^{2}}$. we can see that $\Delta T$ reaches its
maximum value when $\delta _{k}=0$ and $\gamma _{a}=\Gamma
_{R}-\Gamma _{L}$. Fig. 2(b) shows the transmission probabilities
$T_R$, $T_L$ and $\Delta T$ against $\delta_k$ for a two-level
emitter coupled to a 1D waveguide when $\Gamma_L/\Gamma_R=0.1$ and
$\gamma _{a}=\gamma _{R}-\gamma _{L}$.

\begin{figure}[t]
\includegraphics*[width=4cm, height=2.5cm]{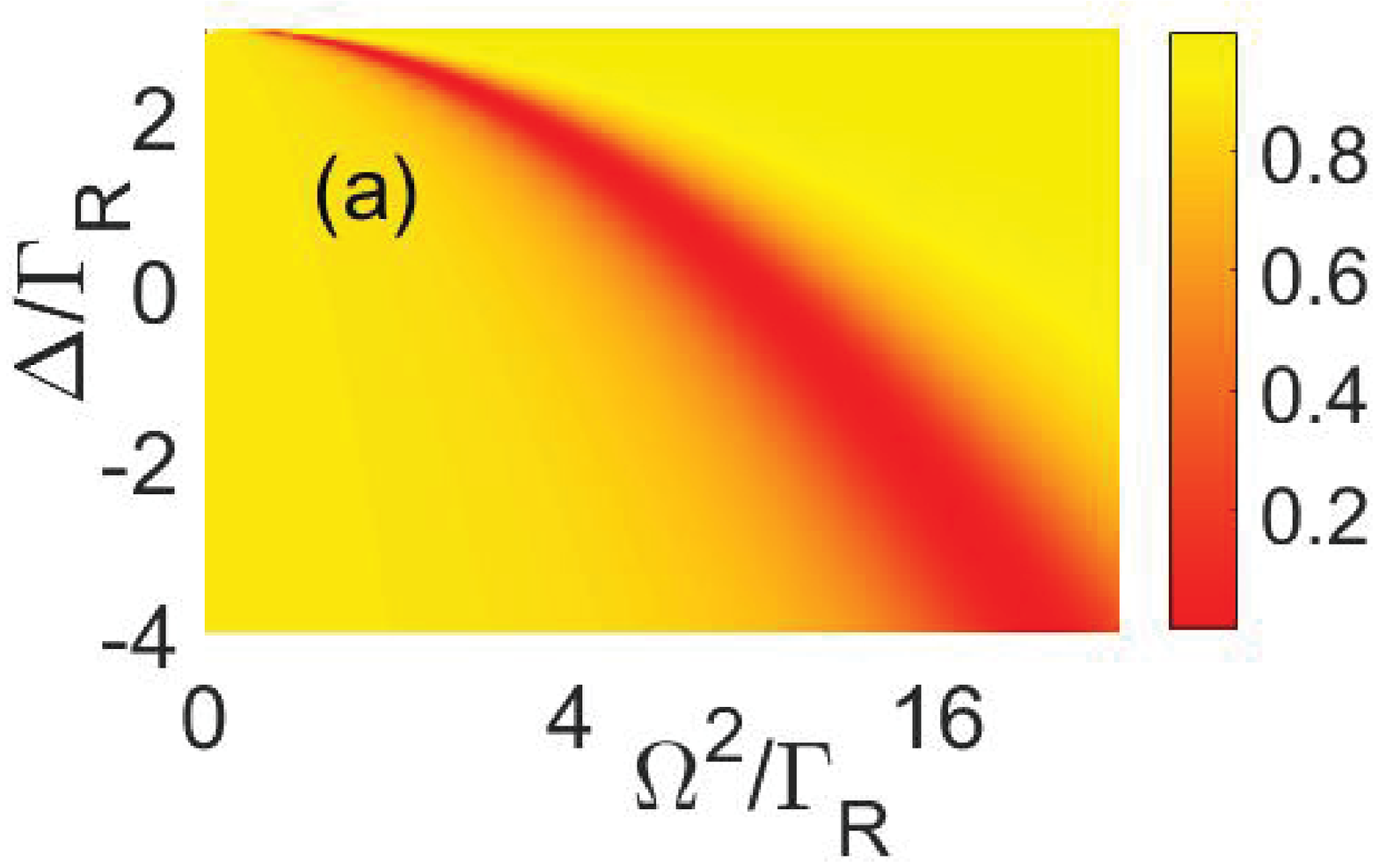} \includegraphics*%
[width=4cm, height=2.5cm]{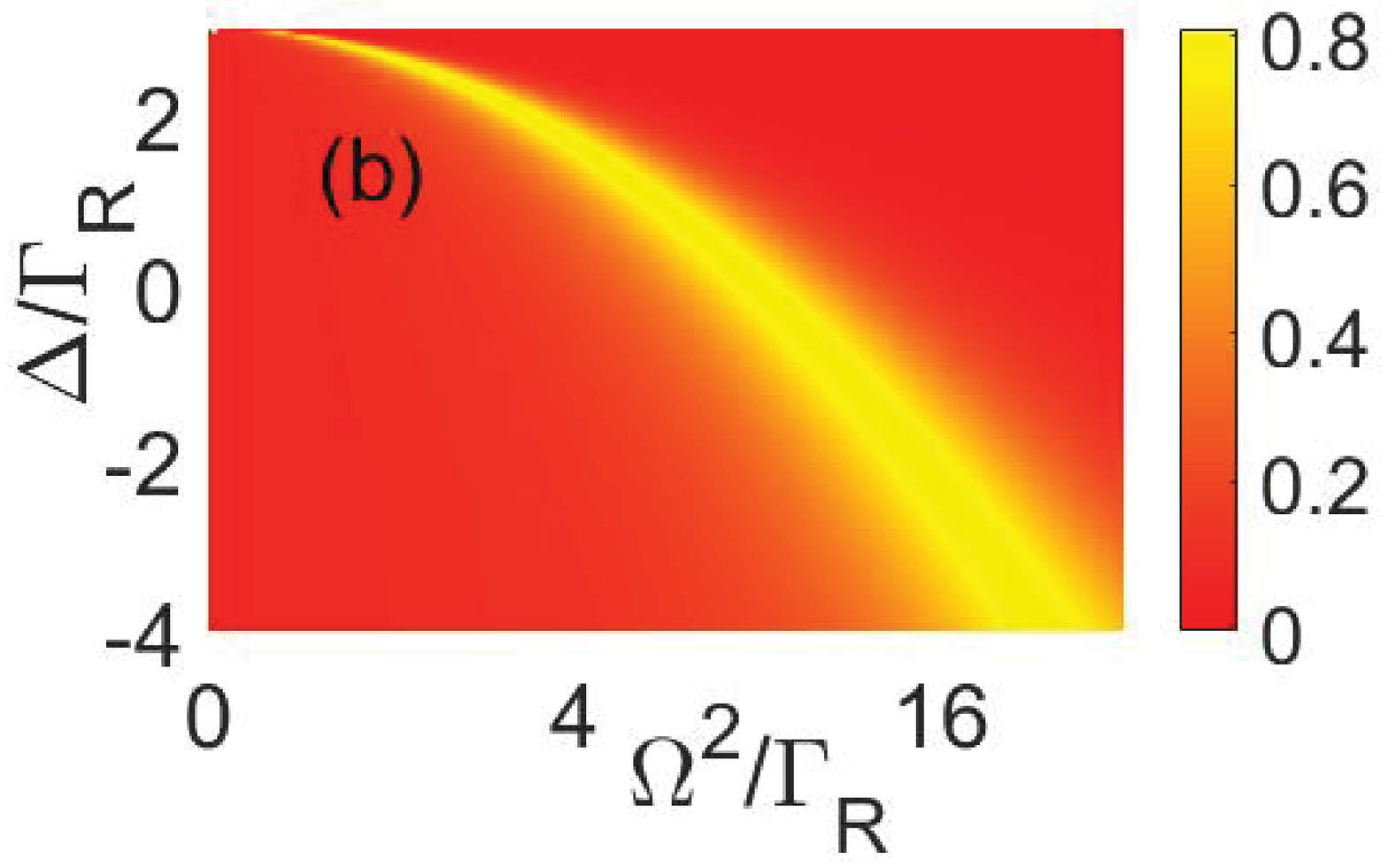} \includegraphics*[width=4cm,
height=2.5cm]{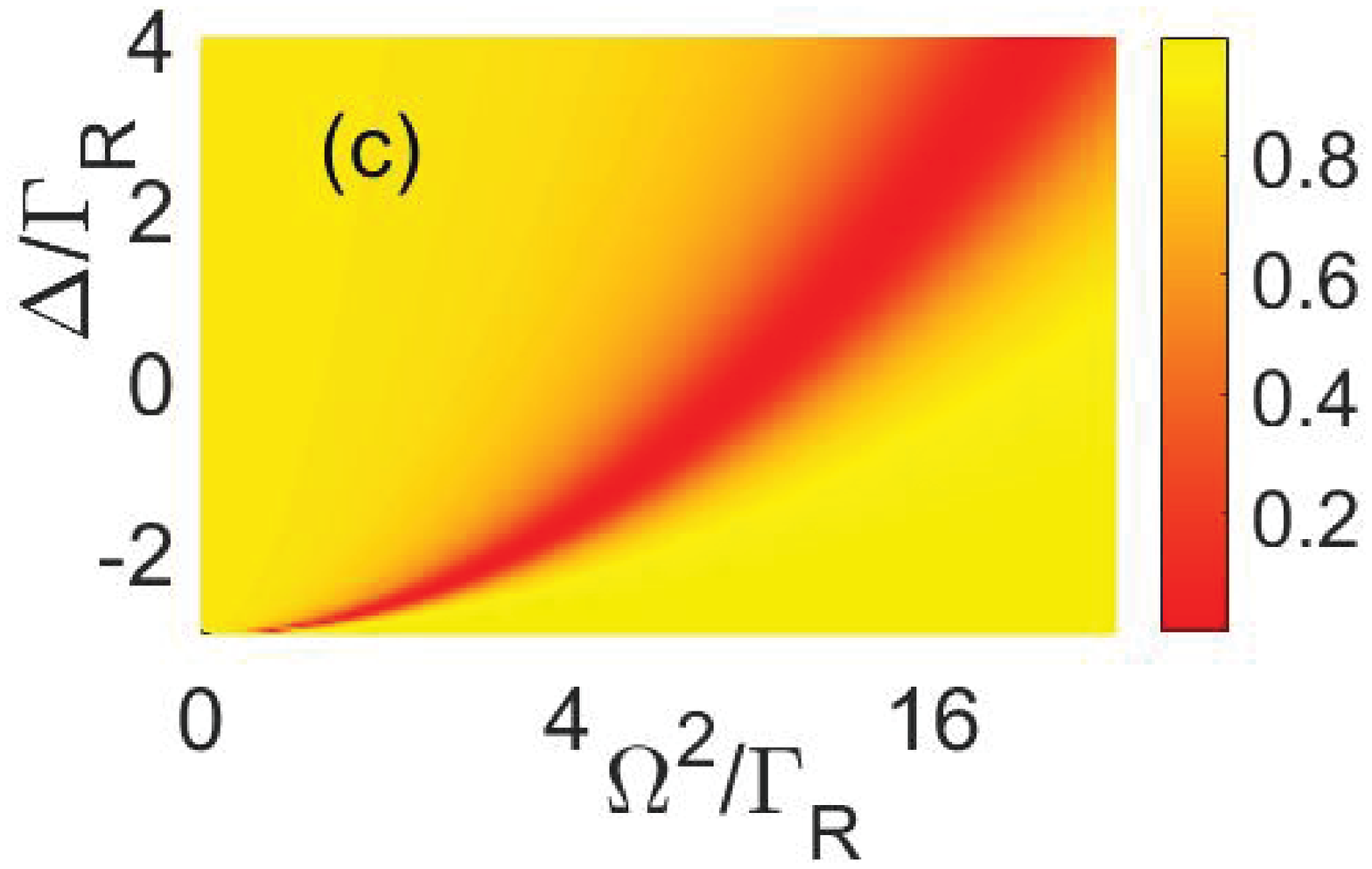} \includegraphics*[width=4cm, height=2.5cm]{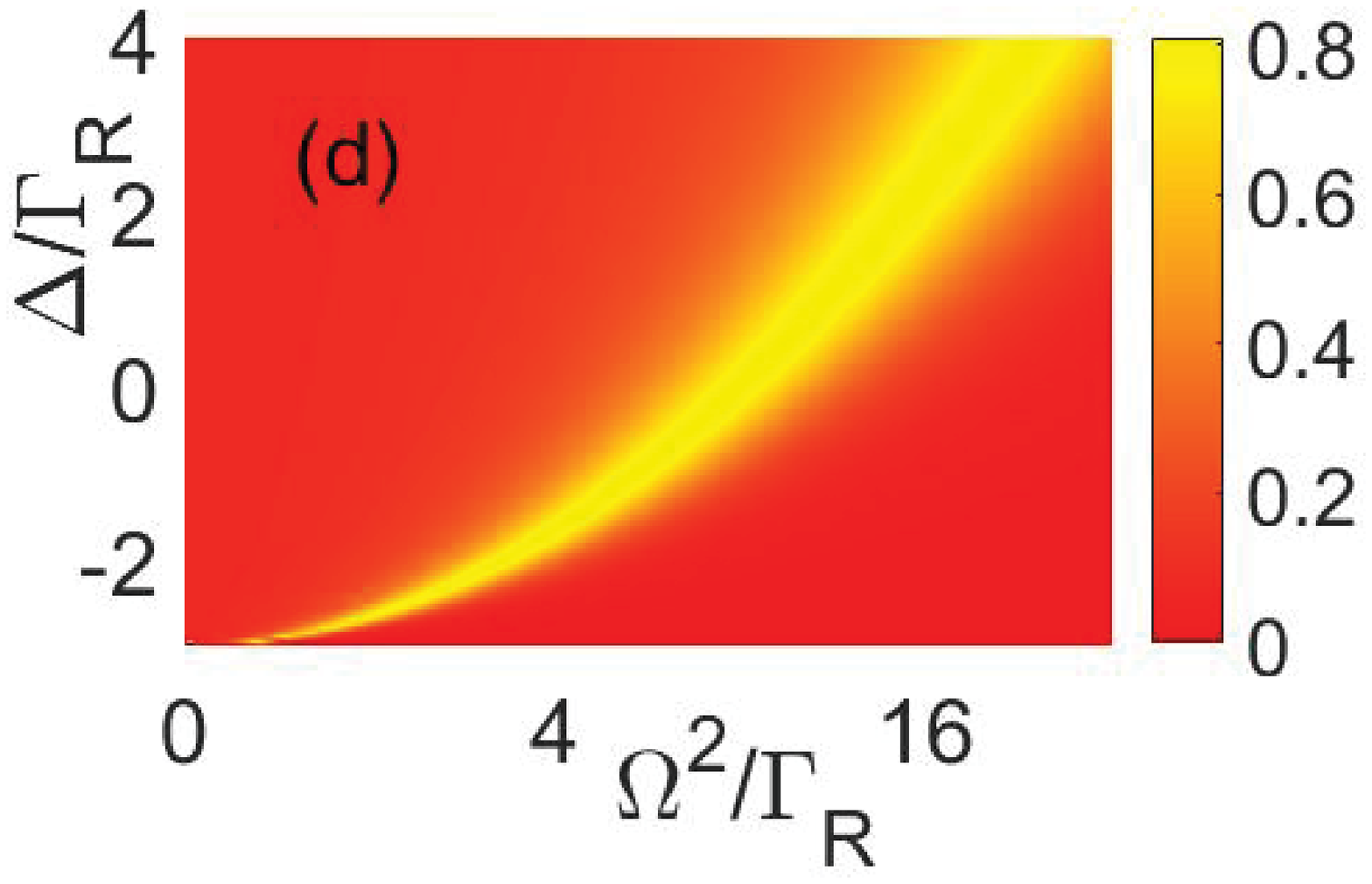%
}
\caption{Single-photon transmission probabilities against the frequency and
Rabi frequency of the external laser. (a) and (c) denote the probability $%
T_R$. (b) and (d) denote $\Delta T$. We take
$\protect\delta_k/\Gamma_R=3$ in (a) and (b),
$\protect\delta_k/\Gamma_R=-3$ in (c) and (d). The other parameters
are: $\Gamma_L/\Gamma_R=0.1$, and $\gamma_a/\Gamma_R=0.9$.}
\end{figure}

When the external laser is turned on, our scheme is a 1D waveguide
chirally coupled to a $\Lambda $-type three-level emitter. In this
case, the single-photon transmission and reflection probabilities
have been obtained in Eqs. (\ref{ampl1}) and (\ref{ampl2}). When
$\Delta _{k}=0$ the single photon transporting towards either
directions will be fully transmitted no matter what the values of
$C$ and $\gamma _{a}$ are due to the interference, which corresponds
to EIT. We plot the single-photon reflection probabilities against
the detuning $\delta _{k}$ when $\gamma _{a}=0$ in Fig. 2(c). The
spectra split due to the interaction between the emitter and the
laser. Similar to the two-level emitter, the symmetrical coupling
reduces the maximum value of $R$. When $\gamma _{a}=\Gamma
_{R}-\Gamma _{L}$ and $\delta
_{k}=\frac{\Delta \pm \sqrt{\Delta ^{2}+4\Omega ^{2}}}{2}$, we obtain $%
T_{R}=0$ and $T_{L}=(\frac{\Gamma _{R}-\Gamma _{L}}{\Gamma _{R}})^{2}$. In
this case, $\Delta T=\frac{\gamma _{a}(\Gamma _{R}-\Gamma _{L})}{\frac{%
(\Delta _{k}\delta _{k}+\Omega ^{2})^{2}}{\Delta
_{k}^{2}}+(\frac{\gamma _{a}+\Gamma _{R}+\Gamma _{L}}{2})^{2}}$\
reaches its maximum value. In Fig. 2(d) we plot the single-photon
transmission probabilities against $\delta _{k}$ for a $\Lambda
$-type three-level emitter. The outcomes provide a manner to realize
the single-photon switch. By adjusting the laser frequency, we can
ensure the single photon is fully transmitted by EIT. Similarly, by
adjusting the frequency and Rabi frequency of the laser, we can
ensure the single photon can not be transmitted by critical
coupling. It is interesting that the emitter's decay is considered
for these operations. The maximum values of $\Delta T$ and $T_L$ are
equal to the corresponding values of the two-level emitter. However,
the $\Lambda $-type three-level emitter provides a control to
various input frequencies.

To ensure the diode works well, the frequency of the single photon
can not be arbitrary. For a two-level emitter, the single-photon
should be nearly resonant to the emitter, i. e., $\delta _{k}\simeq
0$. For a $\Lambda $-type three-level emitter, the single-photon
frequency should satisfy the relation $\delta _{k}=\frac{\Delta \pm
\sqrt{\Delta ^{2}+4\Omega ^{2}}}{2}$. The latter shows the advantage
that it is largely tunable. For various values of the single-photon
frequencies, we can adjust the frequency and Rabi frequency of the
laser to satisfy this condition. Although this condition can not be
satisfied for any arbitrary value of the single-photon frequency, it
can be satisfied in a large range of values. This feasible range is
enough to obtain a single-photon diode which is feasible for various
single-photon frequencies. We plot the probabilities $T_{R}$ and
$\Delta T$ against the laser parameters in Fig. 3. It shows that the
single-photon diode works well at various values of the laser
frequencies and Rabi frequencies.

The single-photon transport scattering by a $\Lambda $-type three-level
emitter can be understood by the dressed-state representation. Our scheme is
considered as the waveguide chirally coupled to a three-level emitter with
states $\left| b\right\rangle $, $\left| +\right\rangle $, $\left|
-\right\rangle $. The transitions between the ground state and dressed
states\ are driven by the photons in the waveguide. The states $\left| \pm
\right\rangle $ are the eigenstates of the Hamiltonian $H_{0}=\omega
_{a}\sigma ^{aa}+(\omega _{c}+\omega _{L})\sigma ^{cc}+\Omega (\sigma
^{ac}+\sigma ^{ca})$, with frequencies $\omega _{a}-\frac{\Delta \mp \sqrt{%
\Delta ^{2}+4\Omega ^{2}}}{2}$. The condition $\delta _{k}=\frac{\Delta \pm
\sqrt{\Delta ^{2}+4\Omega ^{2}}}{2}$ implies that the single photon
resonantly drives one of the transitions $\left| b\right\rangle
\leftrightarrow \left| \pm \right\rangle $.

For two- or three-level emitters coupled to a waveguide, the decay
match condition $\gamma _{a}=|\Gamma _{R}-\Gamma _{L}|$ plays an
important role. The rate $\gamma _{a}$ mainly relates to the
environment surrounding the emitter. The decay rates $\Gamma _{R}\
$and $\Gamma _{L}$ relate to the position of the emitter relative to
the waveguide. Here, the state $\left| c\right\rangle $ is
considered long-live. When $\left| c\right\rangle $ is an excited
state, the dissipation can be incorporated by introducing an extra
nonhermitian term $-i\frac{\gamma _{c}}{2}\sigma ^{cc}$ into the
Hamiltonian
(\ref{hami1}). In addition, the term $\Delta _{k}$ in the results (\ref%
{ampl1}) and (\ref{ampl2}) should be replaced by $\Delta _{k}+i\frac{\gamma
_{c}}{2}$. The tunable single-photon diode can also be achieved in this
case. We will not cover it again.

\section{Conclusions}

We propose a scheme to investigate the tunable single-photon diode. This
diode is composed by an emitter chirally coupled to a 1D waveguide. We study
the single-photon scattering by a two-level and $\Lambda $-type three-level
emitter. The single-photon diode is underpinned by the chirally coupling and
the decay from the emitter's excitation to the other channels except the
waveguide. To make the single-photon diode work well, the frequency of the
single photon must satisfy certain conditions. Especially, for a $\Lambda $%
-type three-level emitter, the diode can work well at various single-photon
frequencies by adjusting the external laser parameters. By tuning the laser
parameters, the single-photon transmission probability can be tuned to zero
or unity. Different from the few-photon diode \cite{nr6roy1}, the
single-photon diode property is not affected by the nonlinear effect in the $%
\Lambda $-type three-level emitter. The $\Lambda $-type emitter in
the tunable single-photon diode can also be replaced by a
three-level emitter in cascade configuration \cite{Yanwb}.

\section{Acknowledgement}

This work is supported by NSFC No. 11505023, No. 11505024, and No. 11774406,
MOST of China No. 2016YFA0302104 and No. 2016YFA0300600, Central of
Excellence in Topological Quantum Computation of Chinese Academy of Sciences
No. XDPB-0803.


\end{document}